\documentclass[a4paper,12pt]{article}
\pdfoutput=1
\usepackage{amsmath}
\usepackage{xcolor}
\usepackage{amsfonts}
\usepackage{amssymb}
\usepackage{amstext}
\usepackage{graphicx}
\usepackage{bm}
\newcommand{\ti}[1]{\ensuremath{ \tilde{#1}}}
\newcommand{\nb}{\ensuremath{ \nabla }}

\newcommand{\m}{\ensuremath{{\mu \nu}}}

\newcommand{\p}{\ensuremath{\partial{}}}

\newcommand{\ot}{\ensuremath{\omega_{\rm tot}}}
\usepackage{epsfig}%
\title{\centerline \bf Ghost condensates and pure kinetic $k$-essence condensates in
presence of field-fluid non-minimal coupling in the dark sector}
\bigskip
\author{Saddam Hussain$^\$$, Anirban Chatterjee$^\star$, Kaushik Bhattacharya$^\ddagger$
\thanks{$^\ddagger$kaushikb@iitk.ac.in, $^\star$anirbanc@iitk.ac.in, $^\$$msaddam@iitk.ac.in}
\\
\normalsize
Department of Physics, Indian Institute of Technology, Kanpur\\ 
\normalsize
Uttar Pradesh 208016, India
}
\begin{document}
\maketitle
\begin{abstract}

In this article we try to find out the conditions when a ghost field in conjunction with a barotropic fluid produces a stable accelerating expansion phase of the universe. It is seen that in many cases the ghost field produces a condensate and drives the fluid energy density to zero in the final accelerating phase, but there can be other possibilities. We have shown that a pure kinetic $k$-essence field (which is not a ghost field) interacting with a fluid can also form an interaction induced condensate and produce a stable accelerating phase of the universe. In the latter case the fluid energy density does not vanish in the stable phase.
\end{abstract}
\section{Introduction}
\label{sec1}
Our present universe is going through an epoch of accelerated expansion. During the last fifteen years, several observational evidences established this fact. Estimation of luminosity distance and redshift of type Ia supernova \cite{Riess:1998cb,Perlmutter:1998np,Riess:2006fw} are the key ingredients among the several observational ventures. Cosmic Microwave Background radiation \cite{Gawiser:2000az}, Baryon acoustic oscillation \cite{Eisenstein:2005su, Percival:2006gs} and Hubble constant \cite{OHD} also play significant roles in the accelerated expansion phase. Literature related to the accelerated expansion of the universe \cite{Riess:1998cb,Perlmutter:1998np,Riess:2006fw} suggests that dark energy  is solely responsible for this late-time cosmic acceleration. The large negative pressure of dark energy prevents gravitational collapse and produces late-time cosmic acceleration.

Despite countless theoretical approaches, the physical theory of dark
energy is still not established. The earliest theory to probe the
nature of dark energy is the $\Lambda$CDM model, which consists of
both cosmological constant $\Lambda$ and cold dark matter. This model
was favored by particle physics community. Unfortunately, the
$\Lambda$CDM model suffers from two drawbacks. The first difficulty of
this model is related to the cosmological constant problem
\cite{Martin:2012bt} and the other one is the cosmic coincidence
problem \cite{Zlatev:1998tr}. These two basic problems motivate us to
study alternative dark energy models. One type of such models involve
the field-theoretic dark energy model, in which the scalar field plays
a major role in producing a negative pressure resulting in the
late-time cosmic acceleration. Based on the form of the Lagrangian, mainly
two kinds of scalar field models exist so far, one is `Quintessence'
model
\cite{Zlatev:1998tr,Peccei:1987mm,Ford:1987de,Peebles:2002gy,Nishioka:1992sg, Ferreira:1997au,Ferreira:1997hj,Caldwell:1997ii,Carroll:1998zi,Copeland:1997et,Hebecker:2000au,Hebecker:2000zb}
and the other is `$k$-essence' model
\cite{ArmendarizPicon:1999rj,Garriga:1999vw,ArmendarizPicon:2000ah,Armendariz-Picon:2000nqq,Chimento:2003zf,Chimento:2003ta,Scherrer:2004au,Chiba:1999ka,Bose:2008ew}. Despite
the presence of scalar field models, other kind of dark energy models
are also present in cosmology, like $f(R)$ theory of gravity
\cite{Capozziello:2002rd,Capozziello:2003gx}, scalar-tensor
theoretical models \cite{Amendola:1999qq} and brane-world models
\cite{Sahni:2002dx}.  In this present work, we will only concentrate
upon some aspects of the $k$-essence type of scalar field dark energy
model. One must note that non-canonical scalar field ($k$-essence type
of scalar field) is not only used to study the nature of dark energy
but they are also regularly used to characterize inflation
\cite{ArmendarizPicon:1999rj,Garriga:1999vw}, dark matter
\cite{Bose:2008ew} and unified dark sector models
\cite{Scherrer:2004au}. Out of the several theoretical ventures to
study the $k$-essence sector, one fascinating method is investigating
the interacting field-fluid framework. Several interlinking
field-fluid scenarios have already been investigated in the literature
\cite{Brown:1992kc,Boehmer:2015kta,Boehmer:2015sha,Shahalam:2017fqt,Barros:2019rdv,Kerachian:2019tar}. In
most of the cases, field part is composed of a Quintessence scalar
field. In some literature, $k$-essence \cite{chatterjee} has taken the
lead role in driving the cosmological dynamics. Algebraic \cite{Boehmer:2015kta}
and derivative \cite{Boehmer:2015sha} types of coupling have been
used to study this interlinked dark sector. Dynamical
stability analysis plays an important role in the study of dark energy. Various applications of dynamical systems are applied in studying evolution
of various field-fluid scenarios.  We have extracted the essence of
dynamical stability technique from various literature
\cite{Boehmer:2015sha,Chakraborty:2019swx,Roy:2017uvr,DeSantiago:2012nk,Dutta:2016bbs,Ng:2001hs,Koivisto:2009fb,Bahamonde:2017ize,Tamanini:2014mpa,Dutta:2017wfd}
and implemented it in our study of the dark sector.

Constant potential $k$-sector defines the purely kinetic $k$-essence
models. The purely kinetic $k$-essence models can unify both the dark
energy and the dark matter \cite{Scherrer:2004au} sectors.
In various dark sector models involving a scalar field and a fluid it is
assumed that the constituents interact gravitationally. Apart from
gravity there may exist some other phenomenological types of interaction
that connect these two sectors. In \cite{chatterjee, Bhattacharya:2022wzu} non-minimal
interaction between the field and the fluid sectors, where the
$k$-essence potential plays a vital role, has been discussed
thoroughly.  Some recent works show that the equation of state of
field-fluid sector of the universe can cross the phantom barrier
\cite{Novosyadlyj:2012vd,Teng:2021cvy,DiValentino:2020naf}. Other
recent works discuss about the issue of Hubble tension. One of the
possible ways to address these issues is related to the introduction
of the non-minimal interaction between field and fluid systems studied
in \cite{Pan:2020zza,Pourtsidou:2013nha,Sharov:2017iue}.  In this
paper we will investigate the field-fluid interaction between a purely
kinetic $k$-essence field sector and the relativistic fluid
sector. This study is important as the behavior of purely kinetic
$k$-essence sector is very different from the standard $k$-essence
sector where the field potential plays an important role. The purely kinetic $k$-essence sector includes the ghost field sector in cosmology. Ghost field cosmology was elegantly introduced in Ref.~\cite{Arkani-Hamed:2003pdi}. All ghost fields are certain forms of purely kinetic $k$-essence fields but all kinetic $k$-essence fields may not be ghost fields. In this paper we discuss the role of ghost fields in cosmology where they are accompanied by a barotropic fluid. 

In this article we have presented the conditions required for producing a stable accelerated phase of expansion of the universe in presence of a ghost field and a barotropic fluid. Most of the time it is seen that the ghost field forms a ghost condensate in the far future when the fluid energy density becomes vanishingly small. It happens for interactions which vanishes when the fluid energy density vanishes. We have also shown a case where the stable accelerating expansion phase is not produced by a ghost condensate. In such a case due to a novel effect induced by field-fluid coupling, a pure kinetic $k$-essence field (which is not a ghost field) forms a condensate and stabilizes the accelerating expansion phase. In the last case the fluid density in the stable phase does not vanish but remains subdominant. In this article we show that a ghost field is not always needed to produce a stable accelerating expansion of the universe. 

The material presented in the article is organized as follows. In
section \ref{k-essence}, we will discuss purely kinetic $k$-essence
theory in a detailed manner and present the connection between kinetic
$k$-essence fields and ghost fields. In section \ref{genout} we
present the main formal results relating to interaction of a ghost
field and a barotropic fluid.  In section \ref{minimal}, we will
introduce a perfect fluid where the fluid and the scalar field do not
exchange energy-momentum between them. In section \ref{non-minimal},
we will introduce the non-minimal coupling and study the field-fluid
theory. Finally, we will conclude in section \ref{conclusion}.
\section{Purely kinetic $k$-essence fields and the ghost connection}
\label{k-essence}

The basic idea of $k$-essence theory was first introduced by Armendariz-Picon et al.\cite{ArmendarizPicon:1999rj, ArmendarizPicon:2000ah} to explain the inflationary scenario of the early Universe; later, this theory has also been applied to study the late time phase of cosmic evolution. The $k$-essence theory is specified by a scalar field potential\footnote{Although we call it a potential, $V(\phi)$, in reality it does not play the role of a potential function in the conventional sense.}, $V(\phi)$, and a non-canonical function of the kinetic term $X \equiv -\frac{1}{2} \nabla_\mu \phi \nabla^\mu \phi$. The action of pure $k$-essence field, minimally coupled to gravity, is given as:
\begin{eqnarray}
S_k&=&\int d^4x \left[\sqrt{-g}\frac{R}{2\kappa^2} -\sqrt{-g}{\mathcal L}(\phi,X)\right]\,.
\label{act}
\end{eqnarray}
Here $\kappa^2 = 8\pi G$. The first term in the square bracket characterizes the Einstein-Hilbert action and second term represents the $k$-essence Lagrangian. We consider a homogeneous and isotropic background given by  Friedmann-Lemaître-Robertson-Walker (FLRW) metric, whose line element is given by:
\begin{equation}
ds^2 = -dt^2 + a(t)^2 d{\bf x}^2\,,
\end{equation}
where $a$ is the scale-factor in the FLRW metric. Variation of the action with respect to $g^{\m}$ produces the energy-momentum tensor:
\begin{eqnarray}
T_{\mu \nu} \equiv -\dfrac{2}{\sqrt{-g}}\dfrac{\delta S}{\delta g^{\mu \nu}}\,.
\label{}
\end{eqnarray}
Hence the energy-momentum tensor of $k$-essence scalar field is 
\begin{eqnarray}
T_{\mu \nu}^{(\phi)} = - {\mathcal L}_{,X}\,(\partial_\mu \phi)(\partial_\nu \phi)-g_{\mu \nu}\,{\mathcal L}\,.
\label{tphi}  
\end{eqnarray}
The energy density and pressure of the $k$-essence field are,
\begin{equation}\label{}
\rho_{\phi} = \mathcal{L} - 2X\mathcal{L}_{,X} \quad \text{and} \quad P_{\phi} = - \mathcal{L}\,.
\end{equation}
The pressure of the $k$-essence field configuration can be chosen as $P_{\phi} = V(\phi)F(X)$. In such a case the energy density will become 
\begin{equation}
\label{rho}
\rho_{\phi} = V(\phi)[2X F_{,X} - F],
\end{equation}
where, $F_{,X} \equiv dF/dX$. We assume that $P_\phi=0$ at $X=0$. This implies that $F(0)=0$. We expect that $\rho_\phi=0$ when $X=0$. For this condition to prevail we require that at $X=0$ one must have $[XF_{,X}]_{X=0}=0$.  Throughout our work we assume $\rho_\phi \ge 0$. 

Unlike the canonical scalar field the $k$-essence scalar field has a negative mass dimension. This makes the kinetic term $X$ dimensionless. The 
mass dimension of the $k$-essence Lagrangian (density), $\mathcal{L}=-V(\phi)F(X)$, solely comes from the potential term $V(\phi)$. The energy momentum tensor of the field is conserved, $\nb_{\mu}T^{\m} = 0$, yielding 
\begin{equation}\label{}
\dot{\rho}_{\phi} = -3H(\rho_{\phi} +P_{\phi})\,.
\end{equation}
where time derivative of any function is expressed as a dot over that function.
Here the Hubble parameter is $H=\dot{a}/a$. The equation of state (EoS) and sound speed in the $k$-essence field sector are given by, $\omega_{\phi} \equiv P_{\phi}/\rho_{\phi}$ and $ c_s^2 = \frac{(\partial p_{\phi} /\partial X)}{(\partial \rho_{\phi}/\partial X)} $. In the present case these quantities are:
\begin{equation}
\label{omega}
\omega_{\phi} = \dfrac{F}{2X F_{,X} - F},\quad c_s^2 = \frac{F_{,X}}{F_{,X} + 2X F_{,XX}},
\end{equation}
where $F_{,XX} \equiv d^2 F/dX^2$.  For a constant potential ($V_0$), varying the $k$-essence action with respect to $\phi$, in the background of the FLRW spacetime, produces the scalar field equation:
\begin{equation}
\label{motion2}
(F_{,X} + 2X F_{,XX})\ddot \phi + 3H F_{,X} \dot \phi = 0\,.
\end{equation}
In the present case the Friedmann equations are
\begin{eqnarray}
3H^2 &=& \kappa^2 \rho_{\phi}\,,
\label{frd1}\\
2\dot{H}+3H^2 &=& -\kappa^2 P_{\phi}\,.
\label{frd2}
\end{eqnarray}
From the Friedmann equations and the purely kinetic $k$-essence scalar field equation it is seen that the scalar field equation has non-trivial critical points (for non-zero values of $X$) when $F_{,X}=0$. The nontrivial solutions have to be non-zero, positive real numbers. In general $X=0$ is also a critical point for the pure kinetic $k$-essence field equation. From the Friedmann equations it is seen that at $X=0$ both $H$ and $\dot{H}$ becomes zero showing an unstable phase\footnote{This conclusion gets modified when a barotropic fluid is also present.}. The trivial critical point $X=0$ is never reached. The nontrivial fixed point solutions are stable when $F_{,XX}>0$ is near the solutions. Near the fixed points, for the pure kinetic $k$-essence field equation one must have $F<0$ so that $\rho_\phi>0$. Near the fixed points we should also have $\omega_\phi=-1$ and $c_s^2=0$. If the fixed points are stable we get  accelerating universe solutions in presence of the field $\phi$. 

The purely kinetic $k$-essence field has a very close relationship with the ghost fields discussed in Ref.~\cite{Arkani-Hamed:2003pdi}. In this reference the authors have elaborately specified the properties of ghost condensates. According to Ref.~\cite{Arkani-Hamed:2003pdi} a ghost field Lagrangian density is given by:
${\mathcal L}=-V_0F(X)$ where
\begin{eqnarray}
F(X)=-X+\chi(X)\,,
\label{ghostk2} 
\end{eqnarray}
where $\chi$ is a function of $X$. The function $\chi$ is such that there exists one (or possibly more) real solution(s) of $F_{,X}=0$. The ghost condensate is formed when $\langle X \rangle \equiv X_c$ where $X_c$ is a particular solution of $F_{,X}=0$. From what has been discussed it becomes clear that all ghost condensate fields are essentially purely kinetic $k$-essence fields but all purely kinetic $k$-essence fields may not form ghost condensate. There can be kinetic $k$-essence fields for which $F(X)$ is not of the form as given in Eq.~(\ref{ghostk2}). There may be theories where $F_{,X}=0$ is never satisfied. In our paradigm these kind of theories are purely kinetic $k$-essence theories which do not have any ghost field correspondence. In 
Ref.~\cite{Arkani-Hamed:2003pdi} the authors have shown that ghost fields can successfully produce a stable accelerating expansion phase of the universe. Our primary attention in this paper is on a stable accelerating phase of the universe in presence of a kinetic $k$-essence field and a barotropic fluid. We will see that the ghost fields in presence of a barotropic fluid can produce an accelerating phase of the universe. It will also be pointed out that sometimes a pure kinetic $k$-essence field, which is not a ghost field, can produce the accelerating phase in presence of a barotropic fluid. In Ref.~\cite{Arkani-Hamed:2003pdi} the authors had worked out the theory of ghost condensates and their interaction with standard model fields in cosmology. 

In the context of purely kinetic $k$-essence Scherrer has shown  \cite{Scherrer:2004au} that both the dark sectors can be unified. In this unification it is seen that, near the stable critical point, the energy-density $\rho_{\rm sch}$ of $k$-essence scalar field is comprised of a constant energy-density and another energy density resembling non-relativistic matter contribution as: 
\begin{equation}\label{schrrer density}
\rho_{\rm sch} \simeq \rho_{0} + \rho_{1} a^{-3}\,.
\end{equation}  
Here $\rho_{0}$ is the constant energy-density representing the cosmological constant and $\rho_{1}$ is coefficient of the matter energy-density near the critical point. In reality what Scherrer had shown was also shown by the authors of Ref.~\cite{Arkani-Hamed:2003pdi}\footnote{Both the papers were published near the same time.}. Scherrer actually worked out the theory of ghost condensates in cosmology.

In this present paper we will try to formulate the cosmology of kinetic $k$-essence field in the presence of a perfect fluid which can interact non-minimally with the field. We think this approach is useful and new and can lead to interesting research in this sector. Before we proceed we want to present some general features of kinetic $k$-essence fields and ghost condensates in presence of a perfect fluid. 
\section{A general outline of the paper}
\label{genout}

In this section we will give a general outline of the work presented in this paper. It contains the main theoretical input of this paper. The rest of the sections will illustrate the validity of the general results. In the present work we will be dealing with the cosmological development of a kinetic $k$-essence field in presence of a barotropic fluid. We will be specially interested in the non-minimal coupling between the field and fluid sectors. The coupling term, introduced in the action\footnote{The non-minimal coupling term in the action will be specified in section \ref{nonminc} (Eq.~(\ref{complact}))introduced later.}, is naturally given by a function of $X$ and the fluid energy-density $\rho$. This coupling gives rise to interaction pressure $P_{\rm int}(\rho,X)$ and interaction energy-density $\rho_{\rm int}(\rho,X)$\footnote{The forms of these functions will become explicit when we will discuss specific models of field-fluid interaction later in section \ref{nonminc}.}. We assume that  
\begin{enumerate}
\item When $P_{\rm int}(\rho,X)$ and $\rho_{\rm int}(\rho,X)$ are nonzero they are functions of $\rho$ and $X$. The functions are such that none of them can be written as a sum of a function of $X$ and a function of $\rho$ at any time. The factors of $\rho$ and $X$ cannot be separated in $P_{\rm int}(\rho,X)$ and $\rho_{\rm int}(\rho,X)$. This assumption is natural because if one can write $P_{\rm int}=p_1(X)+p_2(\rho)+p_3(\rho,X)$ ($p_i$ are some functions) at any instant of cosmic evolution then there appears an ambiguity about the interpretation of the functions $p_i$\footnote{The function $p_3$ does not contain any term which is purely a function of $X$ or $\rho$.}.  Should one interpret $p_1(X)$ as a pure kinetic $k$-essence pressure or as pressure due to interaction? To avoid such ambiguous situations the interaction terms are assumed to be made up of functions in which $X$ and $\rho$ remain additively inseparable.

\item If $\rho\to 0$ then both  $|P_{\rm int}(\rho,X)|$ and $|\rho_{\rm int}(\rho,X)|$ may tend to zero or infinity but cannot take any other values. This assumption is also natural as we do not expect the field and fluid to interact smoothly with each other when the fluid does not exist in the system at all. For those kind of interactions where the interaction terms diverge as  $\rho\to 0$ one never reaches a stable phase where the fluid energy-density vanishes. If an equilibrium is reached, the fluid energy-density always remains finite in that phase.

\item As the interaction terms do not arise from any particular matter sector we assume $\rho_{\rm int}$ can take all possible values. It can be positive, negative or zero. The total energy-density of the system, $\rho_{\rm tot}=\rho + \rho_\phi + \rho_{\rm int} > 0$ where individually $\rho>0$ and $\rho_\phi>0$.  We can sometimes choose the parameters of the theory and the form of the interaction in such a way that $\rho_{\rm int} \ge 0$ throughout the cosmic evolution.
\end{enumerate}
  
We  have specified that if the $\phi$ field is a ghost field then it can
produce a stable accelerating phase of the universe in the far future. The ghost condensate is formed when $F_{,X}(X)=0$ is satisfied for a particular value of $X=X_c$. On the other hand one can also obtain a stable accelerating solution in presence of a pure kinetic $k$-essence field which is not a ghost field in presence of a barotropic fluid. Before we proceed we will like to present some formal discussion on the field-fluid system where the scalar field is a ghost field.   

First we propose a formal statement about the field-fluid system. The statement is as follows: {\it If a pure kinetic $k$-essence field, with the Lagrangian density ${\mathcal L}=-V_0\left[-X+\chi(X)\right]$, in presence of a barotropic fluid with a positive semidefinite EoS forms a ghost condensate in a stable, accelerating, spatially flat FLRW spacetime with $\omega_{\rm tot}=-1$ then the fluid energy-density tends to zero in the far future (when the scale-factor $a$ has increased appreciably from its initial value)}.  

Before we prove the statement we want to clarify certain points. The ghost condensate is formed when $\langle X \rangle \equiv X_c$ as stated earlier. The condensate value is given by $\langle \phi \rangle\equiv \phi_c = c_*t$, where $c_*$ is a constant. In our case the ghost field has an EoS, the fluid has an EoS,  $\omega=P/\rho$, and the field-fluid system has an EoS, $\omega_{\rm tot}$. 

When the field and fluid do not have any coupling, except gravitational coupling, the field and fluid systems evolve separately. In this case in the far future the ghost field will form a condensate, when $P_{\phi,c}=-\rho_{\phi,c}$. Here the subscript $c$ specifies the values of the variables when the stable condensate has formed.
Suppose this state is a stable accelerating solution of the field-fluid system. In that case
$$\omega_{{\rm tot},c}=\frac{P_{\phi,c} + P_c}{\rho_{\phi,c} + \rho_c}=-1\,.$$    
From this equation we get
$$P_{\phi,c} + P_c=-\rho_{\phi,c} - \rho_c\,,$$
or $\omega \rho_c = -\rho_c$. This shows that $\rho_c$ must vanish in the stable accelerating phase when the condensate has formed. 

If the fluid and the field are non-minimally coupled there arises interaction energy-density $\rho_{\rm int}$ and $P_{\rm int}$, where both of these variables are functions of $X$ and $\rho$ in general. If in the asymptotic future a stable  ghost condensate forms then in this case also we must have $P_{\phi,c}=-\rho_{\phi,c}$. In the present case:
$$\omega_{{\rm tot},c}=\frac{P_{\phi,c} + P_c + P_{{\rm int},c}}{\rho_{\phi,c} + \rho_c + \rho_{{\rm int},c}}=-1\,.$$
This equation gives, $P_{\phi,c} + P_c + P_{{\rm int},c}=-\rho_{\phi,c} - \rho_c - \rho_{{\rm int},c}$. From this equation we get
\begin{eqnarray}
\rho_{{\rm int},c}+P_{{\rm int},c}=- \rho_c(1+\omega)\,.
\label{theorem}
\end{eqnarray}
From our general assumptions about the interaction terms we know that
the addition of $\rho_{{\rm int},c}$ and $P_{{\rm int},c}$ must be a
function of $X_c$ and $\rho_c$. In the present case it is seen that
$\rho_{{\rm int},c}+P_{{\rm int},c}$ is independent of $X_c$, this fact
violates our assumption about field-fluid interaction. As a
consequence the above equation can only hold true when $\rho_{{\rm
    int},c}=P_{{\rm int},c}=\rho_c=0$. This implies that in the stable
phase the field-fluid interaction vanishes, and consequently both
$\rho_{{\rm int},c}$ and $P_{{\rm int},c}$ vanishes. The two fluids
become uncoupled and the general statement is satisfied. This ends the
proof.

One may also state the reverse statement: {\it If the barotropic fluid energy-density tends to zero in the far future, in a stable cosmological phase with accelerated expansion, then a pure kinetic $k$-essence field  with the Lagrangian density ${\mathcal L}=-V_0\left[-X+\chi(X)\right]$ in presence of the barotropic fluid, with a positive semidefinite EoS, forms a ghost condensate in a spatially flat FLRW spacetime with $\omega_{\rm tot}=-1$.}

The proof of this statement is as follows. As $\omega>0$, we cannot
have $\omega=-1$ in the stable phase of acceleration. Thus when the
fluids are decoupled the fluid sector energy-density can only
reduce. As a consequence the final state of zero fluid density can in
principle always be reached. When the fluids are not coupled and in
the final stable accelerating state the fluid density goes to zero
then the only agent responsible for the accelerating phase must be the
ghost condensate.

Next we discuss what happens when there is a non-minimal coupling in the field-fluid system. In this case we have $\rho_s=0$ in the asymptotic stable future. Here the subscripts specify stable phase state variables. From our assumptions on field-fluid coupling, specified at the beginning of this section, we know the the interaction terms  $P_{{\rm int},s}$ and $\rho_{{\rm int},s}$ can either be zero or diverge in this phase. If $\rho_{{\rm int},s}$ diverges the total energy density diverges and the final state does not remain stable, a spacetime singularity emerges in this case. Consequently if the final phase has to be stable
one must have  $P_{{\rm int},s}=\rho_{{\rm int},s}=0$. As a consequence in the stable accelerating phase we have only the ghost field and it can produce stable acceleration only when ghost condensate is formed. In such a case, in the final phase we can replace all the subscripts $s$ with $c$ as because the final stable phase also happens to be the phase where the ghost condensate has formed. The proof  of the reverse statement ends here.

From the above statements we can deduce some general results. The general results are as follows:
\begin{enumerate}    

\item In absence of non-minimal coupling a stable accelerating phase of the universe will always be formed when the barotropic fluid (with a positive semidefinite equation of state) density tends to zero and the condensate forms in the far future.  

\item When a stable ghost condensate is formed, in the accelerating phase of the universe, in presence of field-fluid non-minimal coupling the non-minimal coupling term vanishes in the far future and the system becomes decoupled into two noninteracting phases. The fluid density tends to zero in the far future.  

\end{enumerate}
It is seen from the above discussion that if the ghost condensate forms then it reduces the energy-density of the fluid to zero. If dark matter sector is modelled by a fluid then in the stable phase, when the condensate has formed, there will be no remnants of dark matter. On the other hand pure ghost condensates (in absence of any fluid) tend to produce a `matter-like' effect near the stable point, as shown in Eq.~(\ref{schrrer density}). This matter-like part drops out when the stable condensate is formed \cite{Arkani-Hamed:2003pdi}. If the dark sector is really composed of a ghost field one natural question arises: Has the condensate formed? If the ghost field alone is responsible for the dark sector then the answer must be in the negative, as the dark matter density is not zero now. On the other hand if one claims we have reached the stable phase then pure ghost condensate model cannot be the correct model for the dark sector. The quantity which can unravel the nature of the dark sector is the ratio of the dark matter energy-density and the dark energy-density. If this ratio evolves towards zero, as the system stabilizes,  then a ghost sector alone can take care of the dark sector. On the other hand if this ratio tends towards a constant then the ghost field does not remain a viable option.    

In this paper we will show that both of the options, regarding the ratio, are achievable.
The ratio tends to zero when a ghost field is accompanied by a fluid and their interaction vanishes in the far future. On the other hand in presence of  non-minimal interaction which resists the extreme dilution of the fluid energy-density one can actually get a simple pure kinetic $k$-essence configuration which forms a fluid induced condensate. In this case also one obtains a stable accelerating phase where $\langle X \rangle$ is a constant in presence of a constant fluid energy-density. The condensate is not the standard ghost condensate but a fluid induced pure kinetic $k$-essence condensate which we will specify as the $k$-condensate. 
\section {Cosmological dynamics in the presence of a purely kinetic $k$-essence field and a relativistic fluid \label{minimal}}

In this section we will study cosmological dynamics in presence of a kinetic $k$-essence scalar field and a perfect pressure-less fluid. 
In the present case the fluid and the field do not exchange energy-momentum, consequently they do not interact directly. Although the two matter sectors do not interact directly they affect each other gravitationally. In the next section we will deal with the non-minimally coupled field-fluid case. The action for a purely kinetic $k$-essence scalar field and a relativistic \cite{Brown:1992kc} fluid can be written as:
\begin{equation}
S_{\rm min} = S_k + \int d^4x \left[-\sqrt{-g}\rho(n,s) +
J^\mu(\varphi_{,\mu} + s\theta_{,\mu} + \beta_A\alpha^A_{,\mu})\right]\,.
\label{action with fluid}
\end{equation}
In the above action the new term designates the action of a relativistic fluid. The commas in the subscripts specify covariant derivatives, they become partial derivatives for scalar functions. In the fluid action $\rho(n,s)$ denotes the energy-density of the fluid, which depends on particle number density ($n$) and the entropy density per particle ($s$). The variables such as $\varphi, \theta$, and $\beta_A$ are all Lagrange multipliers. The Greek indices run from $0$ to $3$, and $\alpha^A$ is a Lagrangian coordinate of the fluid where $A$ runs from $1 $ to $3$. The current density $J^{\mu}$ is defined as, 
\begin{equation}\label{}
J^\mu = \sqrt{-g}nu^\mu , \quad u^\mu u_\mu=-1 , \quad |J|=\sqrt{-g_{\mu\nu}J^\mu J^\nu}, \quad n=\frac{|J|}{\sqrt{-g}}\,. 
\end{equation}
Here, $u^{\mu}$ is the 4-velocity of the perfect fluid. Variation of the action with respect to $J^{\mu},s, \theta,\varphi, \beta_{A}, \alpha^{A}$ gives rise to constraint equations. In this article we will not require those constraint equations. The only constraint which is worth mentioning is about the constancy of specific entropy density in the cosmological dynamics of the dark sector.    
Variation of $S_{\rm min}$ with respect to $g^{\m}$ yields the energy-momentum tensor for the relativistic fluid as 
\begin{eqnarray}
T_{\mu \nu}^{(M)}= \rho u_{\mu} u_{\nu} + \left(  n\frac{\partial \rho}{\partial n} - \rho \right) (u_{\mu} u_{\nu} + g_{\m})\,,
\end{eqnarray}
which gives us the energy-density of the fluid $\rho $ and pressure $P= \left(  n\dfrac{\partial \rho}{\partial n} - \rho \right) $. In this case the energy-momentum tensor of the field and fluid part is separately conserved i.e,  $ \nb_{\mu}T_M^{\m} = 0 $ and $ \nb_{\mu}T_\phi^{\m} = 0 $. For the purely kinetic case the variation of the action with respect to $\phi$ gives rise to the field equation as given in Eq.~\eqref{motion2}. In the FLRW background the Friedmann equations can be written as,
\begin{eqnarray}
1 &=& \frac{\kappa^2}{3H^2}(\rho_{\phi} + \rho)\,,
\label{eq:FM1}\\
2\dot{H} + 3H^2 &=& -\kappa^2(P_{\phi} + P)\,.
\label{eq:FM2}
\end{eqnarray}
In the next subsection we will explore the dynamics of this field-fluid system using the dynamical stability technique.
\subsection{Dynamical analysis in the case where pure kinetic $k$-essence field and the hydrodynamic fluid interact gravitationally}

To study the dynamical stability of this system  we choose some dimensionless variables as:
\begin{equation}
x = \dot{\phi}\,, \quad z=\frac{H_0}{H}\,,\quad \sigma^2 = \frac{\kappa^2{\rho}}{3H^2}\,, \quad \Omega_{\phi} = \dfrac{\kappa^2 \rho_{\phi}}{3H^2}\,,
\label{eq:M1}
\end{equation}
where $ \sigma^2 $ is related to the fluid energy-density $\rho$ and $ \Omega_{\phi} $ corresponds to $k$-essence energy-density. In defining $z$ we have used the parameter $H_0$. Here $H_0$ is the Hubble parameter at any instant of cosmological time. It is seen that all the five  variables defined above are not independent of each other. Knowing the dynamics of $x$ and $z$ one can evaluate the values of $\sigma^2$ and $\Omega_{\phi}$. Using Eq.~\eqref{eq:FM1} the constrained equation can be expressed as
\begin{eqnarray}
1 =\sigma^2 + \dfrac{\alpha z^2}{3} (x^2 F_{,X} - F)\,,  
\label{eq:M2}
\end{eqnarray}
where the fluid energy-density as $ 0 \le \sigma^2 \le 1 $. Here
\begin{eqnarray}
\alpha \equiv \frac{\kappa^2 V_0}{H_0^2}\,,
\end{eqnarray}
which is a dimensionless constant.  The constraint equation shows that
the phase space dynamics is actually two dimensional. We only require
to investigate the dynamics of $x$ and $z$. From the constraint
equation we can then infer the value of $\sigma$.

The second Friedmann equation can be expressed as:
\begin{eqnarray}
-\frac{\dot{H}}{H^2} &=&  \frac32 \left(\omega \sigma^2 + \dfrac{\kappa^2 P_{\phi}}{3 H^2} + 1\right)\,,
\label{eq:M3}
\end{eqnarray}
where $\omega$ is equation of the state of the fluid. The energy-density and pressure of the scalar field can be written in terms of dynamical variable $ x $ by using the chosen form of  $F(X)$. Till now we have not chosen any particular form of $F(X)$. All the statements made till now, and in most parts of the next section, are in general true for any form of $F(X)$. 

In the present case the set of autonomous equations, in the two dimensional phase space, is given by
\begin{equation}\label{eq:dynxM}
\begin{split}
x'=	\dot{x}/H	& = -\frac{3xF_{,X}}{x^2F_{,XX}+F_{,X}}\,,   \\
z'= \dot{z}/H & = \dfrac{3}{2}z \left[\omega \, \sigma^2 +  \dfrac{\alpha z^2}{3} F+ 1 \right]\,.
\end{split}
\end{equation}
From the definition of $ \Omega_{\phi} $ and $  \mathcal{P}_\phi $, the total (or effective) EoS and adiabatic sound speed $ (c_{s}^2) $ in the $k$-essence sector are written as:
\begin{eqnarray}
\omega_{\rm tot} = \frac{P_{\rm tot}}{\rho_{\rm tot}} = \omega\sigma^2 + \dfrac{\kappa^2 P_{\phi}}{3 H^2} \label{eq:wtotM}\quad 
\text{and}\quad
c_{s}^2 = \frac{dP_{\phi}\big / dX}{d\rho_{\phi}\big / dX} = \frac{F_{,X}}{2XF_{,XX}+F_{,X}}\,,
\label{eq:ctot2}
\end{eqnarray}
where $P_{\rm tot}=P_{\phi} + \omega \rho$ and $\rho_{\rm tot}=\rho_\phi + \rho$.
In the case of fluid, the square of the sound speed is equal to its equation of state $\omega$.

From the autonomous equations it is seen that the stability of the $x$ variable is not directly dependent on $z$. We have $x^\prime=0$ when $F_{,X}=0$ for a nontrivial critical point. Here one can notice that $x=0$ and $z=0$ is always a critical point which is unphysical as $z=0$ demands an extremely high value of the Hubble parameter. In the late phase of the universe one can safely neglect this critical point. When $F_{,X}=0$ one can have $z^\prime=0$ only when $\sigma^2=0$. This can be verified if one uses the constraint relation, Eq.~(\ref{eq:M2}), in the autonomous equation for $z$. The solution is stable if $F_{,XX}>0$. If the last condition is fulfilled we see that in presence of a barotropic fluid the only stable fixed point of the system corresponds to a ghost condensate. In the asymptotic future the fluid energy-density vanishes and the universe settles down to an accelerating phase with $\omega_{\rm tot}=-1$  and $c_s^2=0$. The result is in agreement with the general statements made in the previous section. 

From the above discussion it is clear that near the stable critical point the effect of the fluid is minimal and the system settles down in an accelerated expansion phase purely dictated by the ghost field dynamics. The only effect of the fluid is in the initial transient phase. In the present case the field configuration and the fluid part do not exchange energy-momentum and the stability of the system is solely guided by the ghost field sector. This case is a simple application of our general understanding of cosmological dynamics produced by ghost fields in presence of a barotropic fluid. In the next section we will present two cases of field-fluid interactions where the  cosmological dynamics is relatively more complex. 
\section{Cosmological dynamics in presence of a purely kinetic $k$-essence scalar field non-minimally interacting with a relativistic fluid\label{non-minimal}}
\label{nonminc}

The field-fluid action in the present case corresponds to the action for non-minimal coupling of $k$-essence field and a relativistic fluid \cite{chatterjee}, given by: 
\begin{eqnarray}
S = S_{\rm min}-\int d^4x \sqrt{-g}f(n,s,X)\,.
\label{complact}
\end{eqnarray}
The last term of the action involves the non-minimal coupling term, which is $f(n,s,X)$. The functional form of $f$ depends on the type of interaction. Due to the proposed interaction, some of the fluid equations of motion are now generalized. The non-minimal coupling follows the general properties discussed in section \ref{genout}.

Varying the total action with respect to $g_{\m}$ gives the energy-momentum tensor as
\begin{equation}\label{}
T_{\mu \nu}^{({\rm int})} = n\dfrac{\p f}{\p n} u_{\mu} u_\nu +\left(n\dfrac{\p f}{\p n} -f \right)  g_{\m} - f_{,X}(\p_\mu \phi)(\p_\nu \phi)\,.
\end{equation}
Comparing the above equation with the perfect fluid's energy-momentum tensor yields the energy-density and pressure of the interaction, which are expressed as 
\begin{equation}\label{}
\rho_{\rm int} =f-2X f_{,X}\,, \text{and} \quad P_{\rm int} = \left(n\dfrac{\p f}{\p n} -f \right)\,. 
\end{equation}
Due to the introduction of non-minimal interaction, the energy-momentum tensor of each component of the field-fluid system is not conserved separately; rather, the total energy-momentum tensor is conserved. For more details one can refer to Ref.~\cite{chatterjee}.

We can rewrite the energy-momentum tensors for both sectors as,
$T'_{\m} = T^{(\phi)}_{\mu \nu} - f_{,X}(\p_\mu \phi)(\p_\nu \phi)$
and $\tilde{T}_{\mu \nu} = T^{(M)}_{\m} + T^{({\rm int})}_{\m} +
f_{,X}(\p_\mu \phi)(\p_\nu \phi)$. Hence the Einstein field equation
becomes,
\begin{equation}\label{}
G_{\m} =\kappa^2 (T'_{\mu\nu}+\tilde{T}_{\mu\nu})\,.
\end{equation}
The covariant derivative of the field energy momentum tensor is 
\begin{equation}
\nabla^{\mu} T^{\prime}_{\mu\nu} =  - f_{,X}(\partial_{\mu}\phi) \nabla^{\mu} \nabla_{\nu} \phi \equiv  Q_{\nu}\,.
\label{Conserved scalar field}
\end{equation}
Here, $Q_\nu$ is a 4-vector that defines energy exchange between the systems. Similarly, we can write conservation equation for fluid as,
\begin{equation}\label{}
\nb_{\mu}	\ti{T}^{\m} =  - \dfrac{\p f}{\p X}\nb^{\nu}X \equiv -Q^{\nu}\,,
\end{equation}
showing that the total energy-momentum tensor $\nb^{\mu}(T'_{\mu\nu}+\tilde{T}_{\mu\nu}) = 0$. Varying the action with respect to $\phi$ yields  
\begin{equation}\label{field kessence}
	- 3H\dot{\phi}\left[  \mathcal{L}_{,X} + f_{,X} \right] +  \dfrac{\p }{\p X} (P_{\rm int} +f ) (3H \dot{ \phi})\, - 
	\ddot{ \phi} \left[(\mathcal{L}_{,X} + f_{,X}) + 2X (\mathcal{L}_{,XX} + f_{,XX}) \right]   = 0\,.
\end{equation}
Friedmann equations can be written for this case in the context of FLRW background as, 
\begin{eqnarray}
	3H^2 &=& \kappa^2 \left(\rho_{M}+\rho_{\phi}+\rho_{\rm int}\right)\,,\label{frdnon1}
	\\
	2\dot{H}+3H^2 &=& -\kappa^2 \left(P_{M}+P_{\phi}+P_{\rm int}\right)\,.\label{frdnon2}
\end{eqnarray}
In the next subsection  we will elaborately describe the dynamical technique required to investigate the dynamics of this proposed non-minimally coupled sector. 
\subsection{Dynamical analysis of non-minimally coupled field-fluid scenario}

To analyze the behavior of this non-minimally coupled system we use the dimensionless variables introduced in the last section. The forms of the variables $x$, $\sigma^2$ and $z$ are given in Eq.~(\ref{eq:M1}). To tackle non-minimal field-fluid coupling we introduce some
more dimensionless variables: 
\begin{equation}\label{}
y = \frac{\kappa^2 f}{3H^2}\,,\quad C = \frac{\kappa^2 P_{\rm int}}{3H^2}\; \mbox{and} \quad D =  \frac{\kappa^2 f_{,X}}{3H^2}\,.
\end{equation}
We have introduced $C$ and $D$ as using them we can compactly write the autonomous equations. The constrained equation can be found from Friedmann Eq.~\eqref{frdnon1} and it is:
\begin{equation}\label{constrained eqn nonmin}
1 = \sigma^2 + \dfrac{\alpha z^2}{3} (x^2 F_{,X} - F) + y - x^2 D\,,
\end{equation}
where $ \alpha$ was defined in the previous section. The other Friedmann equation, as given in Eq.~\eqref{frdnon2}, written in terms of the dynamical variables become
\begin{equation}\label{}
\dfrac{2\dot{H}}{3H^2} = - \left( \omega \sigma^2 + \dfrac{\alpha z^2}{3} F + C +1 \right)\,. 
\end{equation}
Here we have used the EoS for the hydrodynamic fluid as $ P = \omega \rho $. The total equation of state of the system is: 
\begin{equation}\label{}
\ot= \dfrac{P_{\rm tot.}}{\rho_{\rm tot}} = \dfrac{\omega \rho + P_{\phi} + P_{\rm int}}{\rho + \rho_{\phi} + \rho_{\rm int}}  = \omega \sigma^2 + \dfrac{\alpha z^2}{3} F + C\,,
\end{equation}
and the sound speed in the $k$-essence sector is: 
\begin{equation}\label{}
c_{s}^2 = \dfrac{P_{\rm tot,X}}{\rho_{\rm tot,X}} = \dfrac{P_{\phi, X} + P_{\rm int ,X}}{\rho_{\phi, X} + \rho_{\rm int ,X}} = \dfrac{ \dfrac{\alpha z^2}{3} F_{,X} + C_{,X}}{ \dfrac{\alpha z^2}{3}(x^2 F_{,XX} + F_{,X}) -x^2 D_{,X}  - D}\,.
\end{equation}
The above sound speed is a suitable generalization of the sound speed in the pure kinetic $k$-essence sector. In order to have a stable theory the sound speed must be positive and satisfy the condition of $ 0 \le c_{s}^2 \le 1 $. Next we explicitly write some of the  cosmological variables which appear in the constraint equation: 
\begin{equation}\label{intomeg}
\Omega_{\phi} \equiv \dfrac{\alpha z^2}{3}(x^2 F_{,X} -F), \quad \text{and} \quad \Omega_{\rm int} \equiv y -x^2 D\,,
\end{equation}
where $ \Omega_{\phi} $, is the energy-density of $k$-essence field and $ \Omega_{\rm int} $ is the interaction energy-density.

Non-minimal interactions make the cosmological system more complex. The first hint of this complexity arises from the form of the interaction energy term $\Omega_{\rm int}$.  For matter components we can always use the standard constraints as $ 0\le \Omega_{\phi}\le 1 $ and $ 0 \le \sigma^2 \le 1 $ respectively. On the other hand $\Omega_{\rm int}$ does not arise from any matter sector and consequently one may have negative values of interaction energy term. The fact that $\Omega_{\rm int}$ is not  positive semi-definite makes the constrained equation Eq.~\eqref{constrained eqn nonmin} less predictive as now both $\Omega_{\phi}$ and $\sigma^2$ can have values greater than one. In the present paper we will try to maintain the conventional bounds on the scalar field energy-density, matter energy-density and interaction energy-density and work out the cosmological dynamics when $ 0\le \Omega_{\phi}\le 1 $, $ 0 \le \sigma^2 \le 1 $ and $ 0\le \Omega_{\rm int}\le 1$ respectively. It must be pointed out here that having $\Omega_{\rm int}=0$, when neither $\rho$ or $X$ is zero, does not imply that non-minimal coupling has vanished, in such cases $P_{\rm int}$ will not be zero. From the results specified in section \ref{genout} we know that in the absence of any spacetime singularity both $P_{\rm int}$ and $\rho_{\rm int}$ are zero when $\rho\to 0$. 

The autonomous equations for the present system are:
\begin{equation}\label{autononminimal01}
\begin{split}
x'=  \dot{x}/H &= \dfrac{ 3x \left( \dfrac{\alpha z^2}{3}  F_{,X} + C_{,X}\right)  }{\left[ (D - \dfrac{\alpha z^2}{3} F_{,X}) + x^2\left( D_{,X} -  \dfrac{\alpha z^2}{3} F_{,XX}\right) \right] } \,, \\
z'= \dot{z}/H & = \dfrac{3}{2}z \left[\omega \, \sigma^2 +  \dfrac{\alpha z^2}{3} F+ C+1 \right].
\end{split}
\end{equation}
This 2-D autonomous system encapsulates the behavior of the system. The variables $ x $ and $ z $ are sufficient to describe the entire behavior of the system.
To proceed further we will choose some form of the interaction term  $ f = \psi(\rho)\xi(X) $, where $\psi(\rho)$ is purely a function of the fluid energy-density $\rho=\rho(n,s)$ and $\xi(X)$ is a function of the kinetic term $X$. As we are dealing with purely kinetic $k$-essence field we assume that the interaction term also to be a function of $\rho$ and $X$. In the whole analysis the field $\phi$ does not appear explicitly in the action. We work with a simple and fairly general form of non-minimal field-fluid interaction, more complicated interaction terms are not ruled out but their analysis will be complicated.  To unravel the nature of field-fluid coupling we will study two cases which will adequately show the rich mathematical structure of these theories. We have made a list of the model parameters in Tab.~[\ref{tab:choices of models}].
In this table $M$ is a parameter with the dimension of mass and $\mathcal{M}= {H_0^2 M^{-4}}/{\kappa^2}$.
\begin{table}[t]
	\centering
	\begin{tabular}{|c |c |c|c|  c|}
		\hline
		$ f $ & $ C $  & $ D $ &  $ y $\\
			\hline
		$g V_0 \rho^{q}
			 X^\beta M^{-4 q} $  & $g [q(\omega+1) -1]  \frac{\alpha z^{2-2q}}{3}\sigma^{2q} 3^{q} \mathcal{M}^{q} X^\beta$ & $\frac{g\alpha z^{2-2q}}{3}\sigma^{2q} 3^{q} \mathcal{M}^{q} \beta X^{\beta-1}$ & $\frac{g\alpha z^{2-2q}}{3}\sigma^{2q} 3^{q} \mathcal{M}^{q} X^\beta $\\
		\hline
	\end{tabular}
	\caption{Model variables where $ f(n,s,X) = \psi(\rho) \xi(X)\,,\,\rho=\rho(n,s)$}
	\label{tab:choices of models}
\end{table}

The form of the interaction term listed above has $f=g V_0 \rho^{q}X^\beta M^{-4 q}$ where $g$ is a dimensionless coupling constant. Here $\beta$ and $q$ are real numbers. The factor $M$ with a mass dimension is present for dimensional reasons. If one chooses $q=1$ then the interaction term becomes $f=g V_0 M^{-4}\rho X^\beta$. The factor $V_0 M^{-4}$ is a dimensionless, positive real constant which can easily be absorbed inside the coupling constant. In such a case the interaction simply becomes $f=g \rho X^\beta$, where we have represented the new, scaled coupling constant by the old symbol $g$.
We will use this kind of interaction to see whether a ghost condensate can actually produce a stable accelerating phase. Later on we will discuss about a model where $q=-1$. 

Before we start analyzing the various cases of non-minimal field-fluid
interaction let us spend some time on explaining an important
difference between these models and the case we dealt previously in
the earlier section. Unlike pure gravitational coupling case, in the
present case the field and fluid sectors directly exchange energy and
momentum with each other and consequently can create or destroy each
other. The scalar field sector can pump energy and momentum to the
fluid sector and can create the fluid or increase or decrease the
energy-density of the pre-existing fluid.  In the present case both
$\rho_\phi$ and $\rho$ can become zero momentarily. In our previous
discussion with noninteracting fluids none of the energy-densities
could become momentarily zero.  There, in the accelerated expansion phase the
fluid energy-density consistently remained zero. In the present case
non-minimal coupling may annihilate one sector momentarily but that
sector again can be created due to the same interaction.  
\subsection{Cosmological dynamics in case I:~$f= g \rho(n,s) X^\beta $}

In the present case we have $q=1$ as a result of which the interaction term is $f= g \rho X^\beta $, where $ g $ is the dimensionless coupling constant. Depending on the dynamics of the system $ g $ can take any real value. This kind of an interaction term becomes exactly zero when $\rho \to 0$. In Tab.~[\ref{tab:choices of models}] some model variables are evaluated that depend on $\beta$ and $\omega$. We can specify some of those variables in the present case as: $ C=g\omega \sigma^2 X^\beta $, $D=g\sigma^2 \beta X^{\beta-1}$ and
$ y=g\sigma^2 X^\beta $. Using these variables and their derivatives with respect to $X$ one can predict the cosmological dynamics in this case.

Before we specify the critical points in this model let us briefly discuss about some of the subtle properties of this non-minimal interaction model. In Model I the interaction energy term is given by:
\begin{equation}\label{}
\Omega_{\rm int} = g \sigma^2 X^\beta(1-2\beta)\,.
\end{equation}
Using the above expression one can write the constraint equation, Eq.~(\ref{constrained eqn nonmin}), as
\begin{equation}\label{const model 1}
1 = \sigma^2\left[1 + g X^\beta(1-2\beta)\right] + \Omega_{\phi}\,,
\end{equation}
where $\Omega_{\phi}$ is a function of $x$ and $z$. From the above equation one can see that if there is a real $X$ for which $1 + g X^\beta(1-2\beta)=0$ then the constraint equation becomes useless at that value (or values) of $X$ as $\sigma^2$ becomes indeterminate at this point (or points). As in the present case the dynamical system used to predict the behavior of the universe is a differential-algebraic system, at those values of $X$ the algebraic component fails resulting in a singularity. The roots of the equation giving rise to singularities are:
\begin{equation}\label{xtobt}
X^\beta = -\dfrac{1}{g(1-2 \beta)}\,.
\end{equation}
From the above expression one can see that in general one can always get 
real values for $X$ and as a consequence the resulting theory is singular. One can avoid such singularities only when
\begin{enumerate}

\item $g<0$ and $\beta>1/2$, or 

\item $g>0$ and $\beta<1/2$.  
\end{enumerate}
The above choice of parameters show that the present non-minimal field-fluid coupling term is constrained. If the parameters $g$ and $\beta$ do not follow the constraints then the cosmological dynamics become indeterminate. Henceforth we will assume that the above constraints are satisfied.

A general discussion on the critical points of the autonomous system of equations given in Eq.~(\ref{autononminimal01}) reveals interesting features. Due to the presence of non-minimal coupling with a barotropic fluid the number of critical points in the present case has increased. We will see shortly that most of the critical points are physically uninteresting. For any arbitrary positive semidefinite $\omega$, it is seen from Eq.~(\ref{autononminimal01}), that there exists a class of critical points of the system for zero fluid energy-density. In this case $C_{,X}=0$ ($C\propto \rho$), the critical points correspond to the critical points in a field-fluid system where the direct coupling vanishes. The critical points are stable if $F_{,XX}>0$ around the solutions of $F_{,X}=0$. Can we have more physically relevant stable critical points, where the fluid density does not vanish? If the EoS of the fluid $\omega>0$ one cannot prove conclusively that such critical points do not exist. This lack of precise understanding of the system does not affect cosmological dynamics, in the dark sector, as we are mainly interested in the field-fluid coupling where the barotropic fluid resembles dark matter with an EoS $\omega=0$. When $\omega=0$ one can conclusively say that the only stable fixed points of the system correspond to the solutions of $F_{,X}=0$. In this case both $C$ and $C_{,X}$ vanishes as they are proportional to $\omega$.  

Except the physically relevant stable critical points of Eq.~(\ref{autononminimal01}), in the present case we can have more critical points due to field-fluid non-minimal coupling. From Eq.~(\ref{autononminimal01}) one can easily see that for $\omega=0$ the system admits a line of critical points at $z=0$. The critical points lie on  the line specified by the points $(x,z=0)$, where one can use any real value of $x$. These set of critical points are physically irrelevant as at these points the Hubble parameter diverges. If the fluid energy-density diverges then the denominator of the first autonomous equation in Eq.~(\ref{autononminimal01}) diverges (as $D\propto \sigma^2$ in the present case) giving rise to $x^\prime=0$. This fact gives rise to another nontrivial critical point when $\omega=0$. This critical point also turns out to be physically irrelevant as we do not expect the fluid energy-density to blow up in an expanding universe in the far future. Both of the critical points discussed in this paragraph are purely mathematical possibilities and it turns out that both of these critical points are unstable. Only one class of stable, physically relevant critical points exist in the present case, for $\omega=0$, and near it the fluid energy-density vanishes. For these class of critical points one can always choose the general form of $F(X)$ as given in Eq.~(\ref{ghostk2}) and consequently the critical points correspond to some form of ghost condensates. This fact was predicted by the general discussion in section \ref{genout}.

Till now we have worked with a general form of $F(X)$. If one wants to predict the dynamical evolution of the system then one must choose some form of $F(X)$ and a specific barotropic fluid. As we are interested in the dark sector we choose the fluid to represent dark matter and consequently we assume $\omega=0$. In this paper we choose the form of $F(X)$ as:
\begin{eqnarray}
F(X)=AX+BX^2\,,
\label{fxform}
\end{eqnarray}
where $A$ and $B$ are non-zero real constants. This form of $F(X)$ satisfies all the properties $F(X)$ requires, as discussed in section \ref{k-essence}. If both $A$ and $B$ are of the same sign then the field $\phi$ is not a ghost field, it is a pure kinetic $k$-essence field. On the other hand if $A$ is negative and $B$ is positive then $\phi$ turns out to be a ghost field (it is still a pure kinetic $k$-essence field). In the present case we see that if we want a stable accelerating phase then $\phi$ must form a ghost condensate, consequently we choose $A=-1$ and $B=1$. With this choice, our $F(X)$ coincides with the form of it as predicted in Eq.~(\ref{ghostk2}) where $\chi(X)=X^2$. With the above choice of $F(X)$ one can easily find out the physically relevant critical point of the autonomous equations in Eq.~(\ref{autononminimal01}). The coordinates of the critical point are $x_c=\pm \sqrt{(-A/B)}$ and $z_c=\pm(2\sqrt{3B/\alpha})/|A|$.
The dynamical solutions in our case is symmetric with respect to the sign of $x$ and consequently we only use the positive value of $x_c$. As we are dealing with an expanding universe we will only consider the case $z_c>0$.
For $A=-1$ and $B=1$ we see that at this critical point $F_{,XX}>0$ and consequently this critical point is a stable critical point.
\begin{figure}[t]
\centering
\includegraphics[scale=.8]{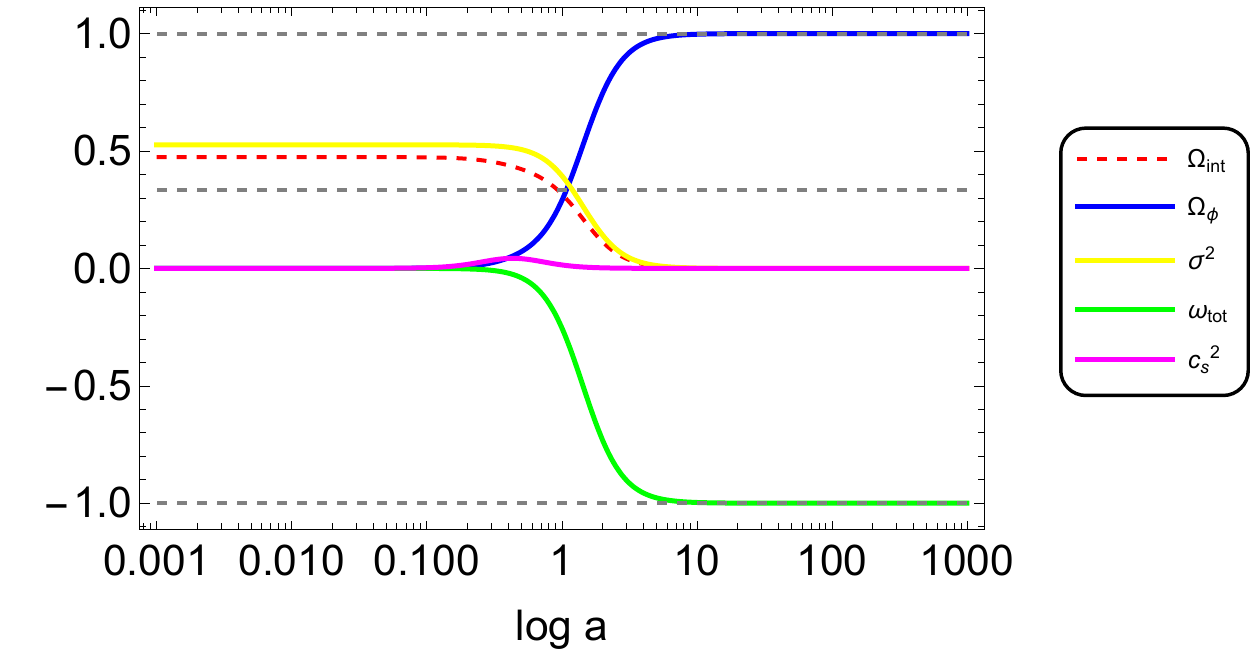}
\caption{Evolution of various Model I variables for $A = -1, B=1, \beta=1/3,\alpha =1,g=3$.}
\label{fig:phase nonmin model1 matter}
\end{figure}

The evolution of the system variables is shown in Fig.~[\ref{fig:phase nonmin model1 matter}]. Here the barotropic fluid sector has an EoS $\omega=0$. From the figure we see that $\Omega_{\rm int}$ starts from a non-zero value and smoothly becomes zero when the ghost condensate forms. Initially $\Omega_{\phi}$ is small but non-zero (a fact not that clear from the curve because of scaling) and remains constant for a while. When  $\Omega_{\rm int}$ starts to dip the ghost field
energy-density parameter, $\Omega_\phi$, smoothly goes up and saturates. The dark matter fluid energy-density parameter, $\sigma^2$, remains constant as long as $\Omega_\phi$ remains constant  and it starts to dip when the ghost condensate starts to form. In the initial phase $\rho/\rho_\phi$ remains a positive constant whose magnitude is greater than one. During this phase the effective system evolves like a matter dominated phase. This resembles some form of scaling behavior which is generally observed in quintessence models of dark energy. In the late phase fluid energy-density tends to zero. The effective fluid EoS, $\omega_{\rm tot}$, initially remains zero and ultimately it settles down to its desired value $-1$, showing a transition from an effective matter dominated phase to an accelerated expansion phase of the universe. The sound speed almost always remains close to zero.

The discussion presented in this section serves as a model for the dark energy dominated universe. It is seen that in the stable accelerating phase, when the ghost condensate is formed, the fluid energy-density tends to zero. The interaction terms vanish in the final phase.  The result turns out to be similar for all non-minimal field-fluid couplings where $\omega=0$. For non-zero EoS the ghost condensate dominated stable point is always present. Can there be models where the dark matter energy-density does not vanish in the stable accelerating phase?  From our general understanding of ghost fields we can say that if the non-minimal interaction is such that it resists the fluid density to vanish in the future then one may obtain a stable accelerating phase with non-zero fluid energy-density. In such a case the scalar field does not have ghost like character, it becomes a purely kinetic $k$-essence condensate which by itself (in the absence of any fluid) is unstable but can form a kinetic $k$-essence condensate only in the presence of a barotropic fluid. In the next section we will present the case that deals with a pure kinetic $k$-essence condensate. 
\subsection{Cosmological dynamics in case II:~$f= g V_0 \rho^{q} X^\beta M^{-4q}$, with $q=-1$}

Here we consider the form of interaction given by $f= g V_0 \rho(n,s)^{q} X^{\beta} M^{-4 q} $. We will specifically deal with the case $q=-1$. The model parameters have been evaluated in Tab.~[\ref{tab:choices of models}]. For a general $q$ the interaction energy term is given by:
\begin{equation}\label{rho_int_md1}
	\Omega_{\rm int} = g \dfrac{\alpha z^{2-2q}}{3} \sigma^{2q}3^{q} \mathcal{M}^{q} X^{\beta}(1-2\beta)\,.
\end{equation}
Using the above expression, one can write the constraint equation as,
\begin{equation}\label{const model 1}
  1 = \sigma^2 + \Omega_{\phi} + \Omega_{\rm int}\,.
\nonumber  
\end{equation}
The theory becomes physically tractable when the field energy-density and fluid density satisfy the conventional constraints $0 \le \Omega_{\phi} \le 1 $, $ 0 \le \sigma^2 \le 1 $. We have worked with parameters which make $\Omega_{\rm int}$ positive semidefinite. For $ q = -1 $ we have  $ f= g V_0 X^{\beta} \dfrac{M^4}{\rho} $. The field component parameter $ (X) $ is not zero in the late-time phases of the universe, but we have seen from our previous discussions that in general a ghost condensate in the final phase tries to diminish the dark matter energy-density down to zero. When the field and fluid had no direct coupling or when the field-fluid coupling vanished in the low matter density regime, formation of ghost condensate in the final phase was a certainty. We know from the general results presented in section \ref{genout} that the interaction terms $|P_{\rm int}|$, and $\rho_{\rm int}$ in the $\rho \to 0$ limit, can either tend to zero or tend to infinity.  
In the previous case both $|P_{\rm int}|$, and $\rho_{\rm int}$ tended to zero as $\rho \to 0$. In the present case we see that both of these variables become unbounded from above in the same limit. This fact shows that if we have a stable phase of accelerated expansion, then in that phase the dark matter energy-density cannot go to zero. The interaction resists the dark matter energy-density to go below a certain threshold and it will be seen that the interaction term also resists ghost condensation.
\begin{table}[t!]
	\centering
	\begin{tabular}{|c|c|c|c|c|c|c|}
		\hline
		Points & $ (x,z) $ & $\Omega_{\phi}$ & $\sigma^2$ & $\omega_{\rm tot}$ & $ c_s^2 $ & Stability\\
		\hline
		\hline
		P  & $ (1.13,1.51) $ & $ 0.55 $ & $ 0.37 $ & $ -1 $ & $ \approx 0  $ & Stable\\
		\hline
	\end{tabular}
	\caption{The physically relevant critical point and its nature corresponding to the parameters $ q=-1, \alpha =1, \beta= 1/3, g=1/2, \mathcal{M}=1, A=1/2,B=1/3 $.}
	\label{tab:critical_model_01}
\end{table}

In the dark matter-dominated phase, when the fluid energy-density dominates, the interaction has  less influence, however, in the final phase, when the fluid energy-density becomes subdominant, interaction becomes stronger between the field and the fluid. In the present case the constraint equation given in Eq.~(\ref{const model 1}) becomes a fourth order algebraic equation in $\sigma$ whose relevant solution for $q=-1$ is given by:
\begin{equation}\label{sigma_mod_1}
	\sigma= \dfrac{\sqrt{1-\Omega_{\phi} + \sqrt{(1- \Omega_{\phi})^2 + 4 (-1 + 2\beta)\tau}}}{\sqrt{2}}\,,
\end{equation}
where $ \tau= g \dfrac{\alpha z^{2-2q}}{3} 3^{q} \mathcal{M}^{q}
X^{\beta} $. Like the previous case in this case also there are some physically irrelevant critical points. For $z=0$ and for any $\omega$ it is seen that all the points on the line $(x,z=0)$ are critical points. These are unstable points. Except these there may be more critical points which lie outside the region of our interest. The choice of model parameters has been made in
such a way that the physically relevant critical points remain real in the region of phase space  constrained by the relations $0 \le \sigma^2 \le 1 $, $ 0 \le \Omega_{\phi} \le 1 $ and $ 0 \le \Omega_{\rm int} \le 1 $. This constrained region of the phase space defines our region of interest. In the present case we assume the barotropic fluid to resemble dark matter and consequently we have $\omega=0$. The form of $F(X)$ was specified in Eq.~(\ref{fxform}). It is seen that there is only one physically relevant critical point corresponding to this model in our region of interest and its properties are tabulated in Tab.~[\ref{tab:critical_model_01}]. The critical point is obtained for a specific set of the model parameters. Changing the model parameters will alter the critical point and for some values of the parameters there may be no critical points. In the present case it is seen that the relevant, stable fixed point is obtained when both $A>0$ and $B>0$.
As a result, in this particular case, it is seen that the stable accelerating phase is obtained only when the scalar field is not a ghost field.

Choosing $ A = 1/2, B = 1/3, \beta= 1/3 $, we have obtained the
relevant critical point of the coupled system numerically.
Numerically one can verify that the fixed point specified in the table is stable.
One must note that in the present case both $A$ and $B$ are positive and
consequently $F(X)_{,X}=0$ cannot be satisfied by any real $X$. This shows that in the present case we are actually dealing with
a pure kinetic $k$-essence field which is not a ghost field. 
\begin{figure}[t!]
	\centering
	\includegraphics[scale=0.7]{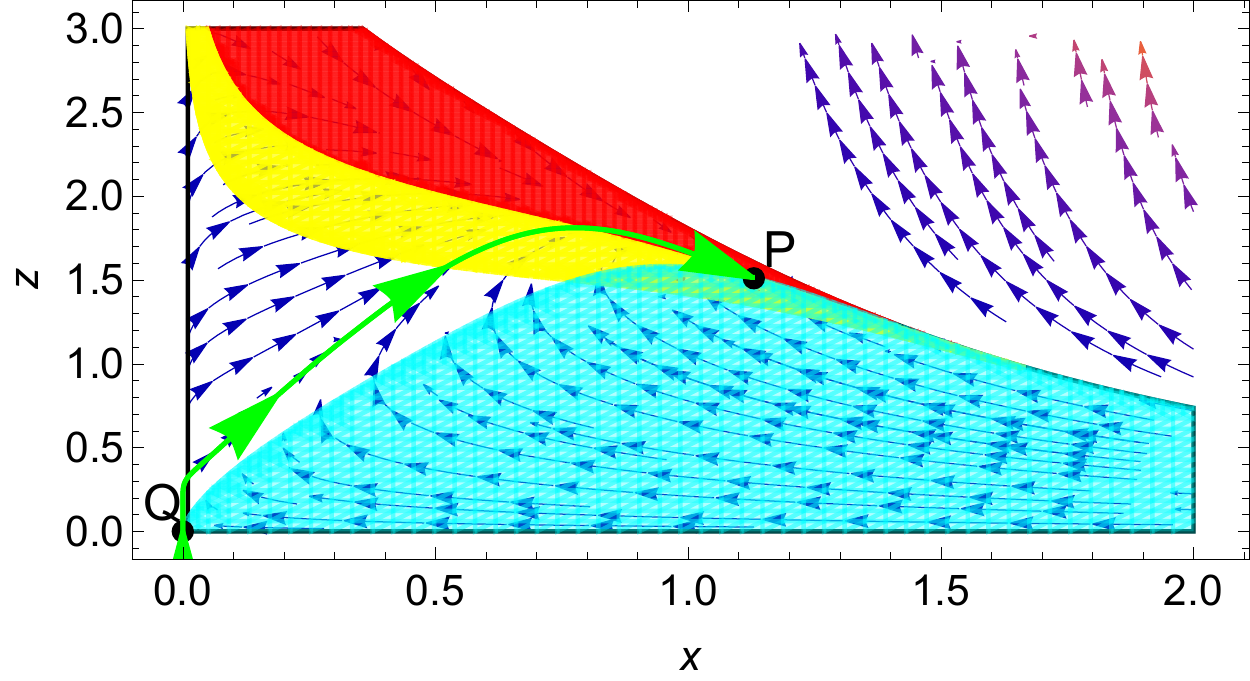}
	\caption{Phase space of non-minimal coupling of $k$-essence with matter fluid for $ A = 1/2, B=1/3, q=-1,\alpha =1, g=1/2, \beta=1/3,\mathcal{M} = 1$.} 
	\label{fig:phase_nonmin_model1_matter}
\end{figure}
The point P, is a stable fixed point specifying accelerated expansion
of the universe. Near this point we have dark energy like behavior
since the $k$-essence field energy-density dominates at this point.

The dynamics near the fixed point can be understood from the phase
space plot in Fig.~[\ref{fig:phase_nonmin_model1_matter}]. The
physically relevant portion of the phase space is coloured and the
shape of this region is dictated by the constraints on the various
parameters $\sigma^2$, $\Omega_\phi$ and $\Omega_{\rm int}$. For a
better understanding of the flow in the phase space we have marked
another reference fixed point Q with coordinates $(0,0)$. A trajectory
evolves from Q and directly goes towards the accelerating point P,
through the non-accelerating region, shown in green. In the phase
portrait the red region specifies that part of phase space where the
system evolves in an effective phantom matter dominated phase, the yellow region
signifies the accelerated expansion phase and in the blue region we
have the sound speed in the scalar field sector to be positive, for
all the points there the sound speed is between $ 0 $ and $ 1
$. Except these there is a white region where none of these conditions
hold.  The arrows that lie outside the constrained region are not
relevant for the present case, they simply show that there can be some
flows in the unconstrained part of the phase space. We remind the reader that there can be other interesting regions in the complete and unconstrained phase space of the system. Due to mathematical complexity of the situation it is very difficult to probe the properties of the whole unconstrained phase space.

\begin{figure}[t!]
	\centering
	\includegraphics[scale=0.8]{"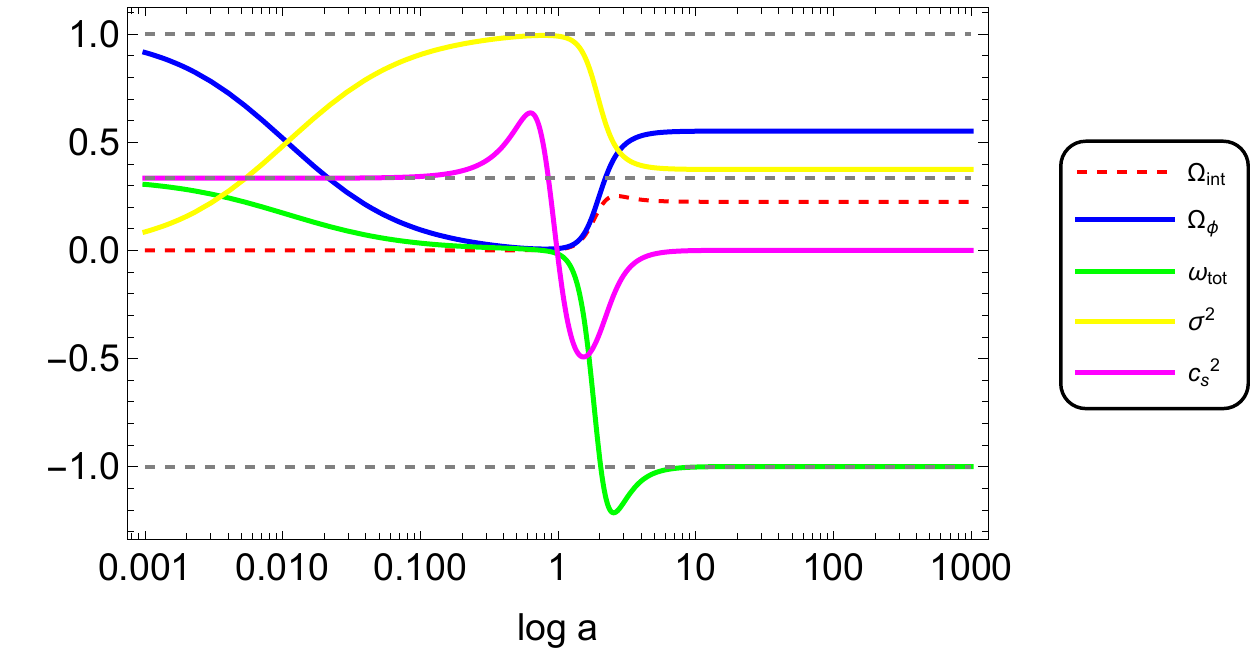"}
	\caption{The evolution plot of various cosmological variables corresponding to $ A = 1/2, B=1/3, q=-1,\alpha =1, g=1/2,\beta=1/3,  \mathcal{M} = 1$. }
	\label{fig:evo_model01_m1}
\end{figure}

We plotted the evolution of various quantities of interest against $\log a $ in Fig.~[\ref{fig:evo_model01_m1}]. In the very early
phase, we see that fluid density $\sigma^{2}$ is less and pure kinetic
$k$-essence energy-density $ \Omega_{\phi} $ is dominating. In this
case, the total equation of state $ \omega_{\rm tot} $ is $ 1/3 $
yielding an effective radiation domination, although there is no
radiation fluid in the system. This demonstrates that the purely kinetic $k$-essence
sector plays an important role as far as the effective EoS of the system is concerned. Over time, the fluid energy-density
grows and eventually overwhelms the field energy-density ushering in
the matter era as the cosmos develops. The EoS is zero (or very nearly
zero) for some period and the speed of sound is seen to be decreasing
in this phase. As the universe further evolves, the kinetic $ k
$-essence sector energy-density increases and so does the interaction
energy-density. The fluid density becomes a non-zero constant in the stable accelerating phase unlike the previous ghost condensate dominated phases. Finally the EoS saturates to $ -1 $, the speed of sound approaches zero symbolizing the presence of dark energy.

In the present case we see that the purely kinetic $k$-essence sector does not correspond to a ghost sector. The purely kinetic $k$-essence sector, in the absence of dust, lacks stability. Only in the presence of non-minimal coupling with dust does $X$ get a time independent expectation value $\langle X \rangle$. As this expectation value is obtained due to non-minimal coupling with a fluid we can call it an induced, kinetic $k$-condensate.  This example shows how the system resists ghost condensate formation in presence of a non-minimal field-fluid coupling which diverges when the fluid density goes to zero. An interesting property of this phase is that the dust energy-density does not diminish as the universe expands. This happens due to the non-zero field-fluid coupling. This case convincingly shows that we may have a stable accelerating phase of the universe even without a ghost condensate.

Before we end this section we will briefly like to specify how these calculations can be connected to real cosmological observations. Primarily our interest was to produce some toy models which specify the cosmological effects of non-minimal coupling between a pure kinetic $k$-essence field and a perfect fluid, which constitute the dark sector in our case. The model involves various parameters such as $V_0, \alpha, g, A, B, q, \beta$ and also the model predictions depend upon the initial conditions we impose upon the autonomous system of equations. This shows that we can actually fit cosmological data by tuning this parameter space. In this paper we have not tried to explain the exact observational predictions, our aim was to show the
qualitative effects of non-minimal field-fluid coupling in cosmology. Nonetheless our work can always be connected to observational cosmology. Here we present briefly some directions in which this can be done. In the future we want to publish a more observation oriented work based on field-fluid coupling.

We can link our work with late-time cosmic observation through two parameters, $viz.$ (i) Coincidence parameter $(r)$, (ii) Deceleration parameter $(q)$. Details about these two parameters are discussed below.
\begin{itemize}
\item Coincidence parameter: The coincidence parameter is defined to be the ratio of dark matter energy density and the dark energy density  \cite{Rabiei:2015pha}  and it can be written as,
\begin{eqnarray}
r = \frac{\rho_{\rm dm}}{\rho_{\rm de}} = \frac{\sigma^2}{\Omega_{\phi}}\Bigg{|}_{\rm our\,\, cases}\,.
\end{eqnarray} 
We can check the coincidence ratio for both of our
  cases and investigate the evolutionary dynamics. From SNe
  Ia+BAO+OHD observation, we can see that $r(z)$ is always a
  decreasing function at late times. From Case I we get initially
  $r(z)$ to be approximately a constant and in Case II initially
  $r(z)$ increases, while in both models $r(z)$ decreases at late
  times. These features can be useful to rule out some of the present
  models. We know from the measurement of satellite borne
  experiments: WMAP \cite{Hinshaw:2008kr} and Planck
  \cite{Ade:2013zuv} that, at present epoch,
  96\% of total energy density of the universe is due to dark matter
  and dark energy, with respective dark matter and dark energy
  contributions amounting to 27\% and 69\% of total energy
  density. This implies that at the present epoch the coincidence ratio is almost
  $0.39$. It is seen that the value of coincidence ratio at late-time is tending to zero for Case I whereas, for Case II the value has the same order of magnitude as that of the observed value. In Case II the ratio we have is more than the observed value by a factor of two but this increase can always be tuned by tuning the parameters of our model.

\item Deceleration parameter: Deceleration parameter \cite{Rabiei:2015pha} can be expressed in terms of Hubble's parameter as,
\begin{eqnarray}
q = -1-\frac{\dot{H}}{H^2}\,.
\end{eqnarray}
In the late-time accelerating cosmological scenarios, the value of $q$  is always negative at different values of redshift. If in the future  an attempt is made to match the results of our calculations with actual observational data then one can 
compute the evolution of $q$ from the early to late-time phase. As time evolves $q$ changes its sign from positive to negative according to SNe Ia observation. One can in principle figure out the nature of evolution of the deceleration parameter for both these models and check our result with the recent observational work \cite{Camarena:2019moy}. Near the critical point the value of  deceleration parameter ($q$) is always tending to -1 for both the cases we studied.
\end{itemize}
\section{Conclusion\label{conclusion}}

In this paper we have shown the link between pure kinetic $k$-essence fields and ghost fields in cosmology. The linkage is not new. All ghost fields are pure kinetic $k$-essence fields whereas all pure kinetic $k$-essence fields may not be ghost fields. The present paper probes a particular feature about the fate of the universe in presence of a ghost field and a barotropic fluid. It is seen that in isolation, a ghost field in cosmology, will in general always lead to a stable accelerating phase of the universe. Does it do so even in the presence of a barotropic fluid, even when a non-minimal coupling is present between the field and the fluid?

We tried to answer the above question formally and our result showed that a ghost field in presence of dark matter, modelled as a fluid with $\omega=0$, will in general form a ghost condensate at late times and lead to a stable accelerating phase. The ghost sector can answer most of the things related to the dark sector, but a question remains. It is seen that in the stable phase, when the condensate has formed, there will be no trace of matter. If one claims that the stable phase has already been attained, as $\omega_{\rm tot}$ is approximately near $-1$ today (or is assumed to be near it as various observations predict its value near $-1$), then one faces a difficulty. The dark matter energy-density today is not zero. On the other hand if one claims that we have yet not reached the stable point then the effective EoS should not be near $-1$. Our general results predict that if there are interactions which forbid the fluid energy-density to vanish in the future then the ghost condensate will not form and a stable accelerated expansion phase can be achieved where the fluid energy-density remains a subdominant constant. Consequently, if the universe has entered the stable phase then our model shows that it can retain a subdominant matter energy-density. The unique signature of this kind of non-minimal interaction is that the matter energy-density remains a constant and does not scale with the $a^{-3}$ factor. 

Our work verified the general results in various cases using a particular form of kinetic $k$-essence Lagrangian density. When the field and fluid are not directly interacting a ghost condensate always forms in the future when the fluid density vanishes. In presence of non-minimal interaction between the scalar field and the fluid, with an arbitrary positive semidefinite EoS, the formation of ghost condensate is always a possibility, in the stable accelerated expansion phase when the interaction term is such that it vanishes when the fluid energy-density vanishes. In particular, when the fluid corresponds to dark matter whose EoS $\omega=0$, the formation of ghost condensate in the accelerated expansion phase is a certainty when the interaction term satisfies the above-mentioned property.

We have shown there can be field-fluid interactions which forbid ghost condensate in the stable phase even when the fluid corresponds to the dark matter sector with $\omega=0$. Such interactions tend to diverge when the fluid energy-density tends to zero. Although we have verified the general predictions made in this paper using a particular form of pure kinetic $k$-essence Lagrangian density and some typical forms of field-fluid interactions, the general predictions hold for all complicated kinds of $F(X)$ 
and all complex forms of field-fluid interactions which satisfy certain basic properties.    

We want to emphasize that we have found a novel phase of pure kinetic $k$-essence condensate which only forms in the presence of a barotropic fluid when the field and fluid sectors are interacting non-minimally. We have shown that there can be a pure kinetic
$k$-essence sector which has no stability in isolation but attains stability only when this sector interacts with a barotropic fluid. One can think of the stable value of $X$ attained in this phase as a condensate, but it is an interaction induced condensate.  

In conclusion we want to state that, most of the time ghost fields and a barotropic fluid can produce a stable accelerating phase when the ghost condensate is formed but not always. There can be cases where pure kinetic $k$-essence fields, which are not ghost fields, can also produce a stable accelerating expansion phase of the universe where matter energy-density remains subdominant.  
\paragraph{Acknowledgement}
A.C would like to thank Indian Institute of Technology, Kanpur for supporting this work by means of Institute Post-Doctoral Fellowship \textbf{(Ref.No.DF/PDF197/2020-IITK/970)}.



\begin{thebibliography}{100}
	
	\bibitem{Riess:1998cb}
	A.~G.~Riess \textit{et al.} [Supernova Search Team],
	Astron. J. \textbf{116} (1998), 1009-1038
	doi:10.1086/300499
	[arXiv:astro-ph/9805201 [astro-ph]]
	
	
	\bibitem{Perlmutter:1998np}
	S.~Perlmutter \textit{et al.} [Supernova Cosmology Project],
	Astrophys. J. \textbf{517} (1999), 565-586
	doi:10.1086/307221
	[arXiv:astro-ph/9812133 [astro-ph]].
	
	\bibitem{Riess:2006fw}
	A.~G.~Riess, L.~G.~Strolger, S.~Casertano, H.~C.~Ferguson, B.~Mobasher, B.~Gold, P.~J.~Challis, A.~V.~Filippenko, S.~Jha and W.~Li, \textit{et al.}
	Astrophys. J. \textbf{659} (2007), 98-121
	doi:10.1086/510378 [arXiv:astro-ph/0611572 [astro-ph]]
	
	
	\bibitem{Gawiser:2000az}
	E.~Gawiser and J.~Silk,
	Phys. Rept. \textbf{333} (2000), 245-267
	doi:10.1016/S0370-1573(00)00025-9
	[arXiv:astro-ph/0002044 [astro-ph]].
	
	
	\bibitem{Eisenstein:2005su}
	D.~J.~Eisenstein \textit{et al.} [SDSS],
	Astrophys. J. \textbf{633} (2005), 560-574
	doi:10.1086/466512
	[arXiv:astro-ph/0501171 [astro-ph]].
	
	\bibitem{Percival:2006gs}
	W.~J.~Percival, R.~C.~Nichol, D.~J.~Eisenstein, D.~H.~Weinberg, M.~Fukugita, A.~C.~Pope, D.~P.~Schneider, A.~S.~Szalay, M.~S.~Vogeley and I.~Zehavi, \textit{et al.}
	Astrophys. J. \textbf{657} (2007), 51-55 doi:10.1086/510772
	[arXiv:astro-ph/0608635 [astro-ph]].
	
	
	
	
	\bibitem{OHD}
	A.~G.~Riess, et al.,Astrophys. J.\textbf{699}(2009) 539 
	[arXiv:0905.0695 [astro-ph.CO]].
	
	
	\bibitem{Martin:2012bt}
	J.~Martin,
	Comptes Rendus Physique \textbf{13} (2012), 566-665
	doi:10.1016/j.crhy.2012.04.008
	[arXiv:1205.3365 [astro-ph.CO]].
	
	
	\bibitem{Zlatev:1998tr}
	I.~Zlatev, L.~M.~Wang and P.~J.~Steinhardt,
	Phys. Rev. Lett. \textbf{82} (1999), 896-899
	doi:10.1103/PhysRevLett.82.896
	[arXiv:astro-ph/9807002 [astro-ph]].
	
	
	
	\bibitem{Peccei:1987mm}
	R.~D.~Peccei, J.~Sola and C.~Wetterich,
	Phys. Lett. B \textbf{195} (1987), 183-190
	doi:10.1016/0370-2693(87)91191-9
	
	\bibitem{Ford:1987de}
	L.~H.~Ford,
	Phys. Rev. D \textbf{35} (1987), 2339
	doi:10.1103/PhysRevD.35.2339
	
	\bibitem{Peebles:2002gy}
	P.~J.~E.~Peebles and B.~Ratra,
	Rev. Mod. Phys. \textbf{75} (2003), 559-606
	doi:10.1103/RevModPhys.75.559
	[arXiv:astro-ph/0207347 [astro-ph]].
	
	\bibitem{Nishioka:1992sg}
	T.~Nishioka and Y.~Fujii,
	Phys. Rev. D \textbf{45} (1992), 2140-2143
	doi:10.1103/PhysRevD.45.2140
	
	\bibitem{Ferreira:1997au}
	P.~G.~Ferreira and M.~Joyce,
	Phys. Rev. Lett. \textbf{79} (1997), 4740-4743
	doi:10.1103/PhysRevLett.79.4740
	[arXiv:astro-ph/9707286 [astro-ph]].
	
	\bibitem{Ferreira:1997hj}
	P.~G.~Ferreira and M.~Joyce,
	Phys. Rev. D \textbf{58} (1998), 023503
	doi:10.1103/PhysRevD.58.023503
	[arXiv:astro-ph/9711102 [astro-ph]]
	
	\bibitem{Caldwell:1997ii}
	R.~R.~Caldwell, R.~Dave and P.~J.~Steinhardt,
	Phys. Rev. Lett. \textbf{80} (1998), 1582-1585
	doi:10.1103/PhysRevLett.80.1582
	[arXiv:astro-ph/9708069 [astro-ph]].
	
	\bibitem{Carroll:1998zi}
	S.~M.~Carroll,
	Phys. Rev. Lett. \textbf{81} (1998), 3067-3070
	doi:10.1103/PhysRevLett.81.3067
	[arXiv:astro-ph/9806099 [astro-ph]]
	
	\bibitem{Copeland:1997et}
	E.~J.~Copeland, A.~R.~Liddle and D.~Wands,
	Phys. Rev. D \textbf{57} (1998), 4686-4690
	doi:10.1103/PhysRevD.57.4686
	[arXiv:gr-qc/9711068 [gr-qc]].
	
	
	\bibitem{Hebecker:2000au}
	A.~Hebecker and C.~Wetterich,
	Phys. Rev. Lett. \textbf{85} (2000), 3339-3342
	doi:10.1103/PhysRevLett.85.3339
	[arXiv:hep-ph/0003287 [hep-ph]].
	
	\bibitem{Hebecker:2000zb}
	A.~Hebecker and C.~Wetterich,
	Phys. Lett. B \textbf{497} (2001), 281-288
	doi:10.1016/S0370-2693(00)01339-3
	[arXiv:hep-ph/0008205 [hep-ph]]		
	
	\bibitem{ArmendarizPicon:1999rj}
	C.~Armendariz-Picon, T.~Damour and V.~F.~Mukhanov,
	Phys. Lett. B \textbf{458} (1999), 209-218
	doi:10.1016/S0370-2693(99)00603-6
	[arXiv:hep-th/9904075 [hep-th]]
	
	
	\bibitem{Garriga:1999vw}
	J.~Garriga and V.~F.~Mukhanov,
	Phys. Lett. B \textbf{458}, 219-225 (1999)
	doi:10.1016/S0370-2693(99)00602-4
	[arXiv:hep-th/9904176 [hep-th]].
	
	\bibitem{ArmendarizPicon:2000ah}
	C.~Armendariz-Picon, V.~F.~Mukhanov and P.~J.~Steinhardt,
	Phys. Rev. D \textbf{63} (2001), 103510
	doi:10.1103/PhysRevD.63.103510
	[arXiv:astro-ph/0006373 [astro-ph]]
	
	\bibitem{Armendariz-Picon:2000nqq}
	C.~Armendariz-Picon, V.~F.~Mukhanov and P.~J.~Steinhardt,
	Phys. Rev. Lett. \textbf{85} (2000), 4438-4441
	doi:10.1103/PhysRevLett.85.4438
	[arXiv:astro-ph/0004134 [astro-ph]].	
	
	\bibitem{Chimento:2003zf}
	L.~P.~Chimento and A.~Feinstein,
	Mod. Phys. Lett. A \textbf{19} (2004), 761-768
	doi:10.1142/S0217732304013507
	[arXiv:astro-ph/0305007 [astro-ph]]	
	
	\bibitem{Chimento:2003ta}
	L.~P.~Chimento,
	Phys. Rev. D \textbf{69} (2004), 123517
	doi:10.1103/PhysRevD.69.123517
	[arXiv:astro-ph/0311613 [astro-ph]].	
	
\bibitem{Scherrer:2004au}
	R.J.Scherrer,
	Phys.Rev.Lett. \textbf{93}, 011301 (2004)
	doi:10.1103/PhysRevLett.93.011301
	[arXiv:astro-ph/0402316 [astro-ph]].
	
	
	\bibitem{Chiba:1999ka}
	T.~Chiba, T.~Okabe and M.~Yamaguchi,
	Phys. Rev. D \textbf{62} (2000), 023511
	doi:10.1103/PhysRevD.62.023511
	[arXiv:astro-ph/9912463 [astro-ph]].
	
	
	
	
	
	\bibitem{Bose:2008ew}
	N.~Bose and A.~S.~Majumdar,
	Phys. Rev. D \textbf{79} (2009), 103517
	doi:10.1103/PhysRevD.79.103517
	[arXiv:0812.4131 [astro-ph]].			
	
	
	\bibitem{Capozziello:2002rd}
	S.~Capozziello,
	Int. J. Mod. Phys. D \textbf{11} (2002), 483-492
	doi:10.1142/S0218271802002025
	[arXiv:gr-qc/0201033 [gr-qc]].
	
	\bibitem{Capozziello:2003gx}
	S.~Capozziello, V.~F.~Cardone, S.~Carloni and A.~Troisi,
	Int. J. Mod. Phys. D \textbf{12} (2003), 1969-1982
	doi:10.1142/S0218271803004407
	[arXiv:astro-ph/0307018 [astro-ph]].
	
	
	\bibitem{Amendola:1999qq}
	L.~Amendola,
	Phys. Rev. D \textbf{60} (1999), 043501
	doi:10.1103/PhysRevD.60.043501
	[arXiv:astro-ph/9904120 [astro-ph]].
	
	\bibitem{Sahni:2002dx}
	V.~Sahni and Y.~Shtanov,
	JCAP \textbf{11} (2003), 014
	doi:10.1088/1475-7516/2003/11/014
	[arXiv:astro-ph/0202346 [astro-ph]]
	
	
	
	\bibitem{Brown:1992kc}
	J.~D.~Brown,
	Class. Quant. Grav. \textbf{10} (1993), 1579-1606
	doi:10.1088/0264-9381/10/8/017
	[arXiv:gr-qc/9304026 [gr-qc]].
	
	\bibitem{Boehmer:2015kta}
	C.~G.~Boehmer, N.~Tamanini and M.~Wright,
	Phys. Rev. D \textbf{91} (2015) no.12, 123002
	doi:10.1103/PhysRevD.91.123002
	[arXiv:1501.06540 [gr-qc]].
	
	\bibitem{Boehmer:2015sha}
	C.~G.~Boehmer, N.~Tamanini and M.~Wright,
	Phys. Rev. D \textbf{91} (2015) no.12, 123003
	doi:10.1103/PhysRevD.91.123003
	[arXiv:1502.04030 [gr-qc]].
	
	
	\bibitem{Shahalam:2017fqt}
	M.~Shahalam, S.~D.~Pathak, S.~Li, R.~Myrzakulov and A.~Wang,
	Eur. Phys. J. C \textbf{77} (2017) no.10, 686
	doi:10.1140/epjc/s10052-017-5255-1
	[arXiv:1702.04720 [gr-qc]].
	
	\bibitem{Barros:2019rdv}
	B.~J.~Barros,
	Phys. Rev. D \textbf{99} (2019) no.6, 064051
	doi:10.1103/PhysRevD.99.064051
	[arXiv:1901.03972 [gr-qc]].
	
	\bibitem{Kerachian:2019tar}
	M.~Kerachian, G.~Acquaviva and G.~Lukes-Gerakopoulos,
	Phys. Rev. D \textbf{99} (2019) no.12, 123516
	doi:10.1103/PhysRevD.99.123516
	[arXiv:1905.08512 [gr-qc]].
	
	\bibitem{chatterjee}
	A.~Chatterjee, S.~Hussain and K.~Bhattacharya,
	Phys. Rev. D \textbf{104} (2021) no.10, 2021
	doi:10.1103/PhysRevD.104.103505
	[arXiv:2105.00361 [gr-qc]].


\bibitem{Bhattacharya:2022wzu}
K.~Bhattacharya, A.~Chatterjee and S.~Hussain,
[arXiv:2206.12398 [gr-qc]].        
	
	
	\bibitem{Chakraborty:2019swx}
	A.~Chakraborty, A.~Ghosh and N.~Banerjee,
	Phys. Rev. D \textbf{99} (2019) no.10, 103513
	doi:10.1103/PhysRevD.99.103513
	[arXiv:1904.10149 [gr-qc]].
	
	\bibitem{Roy:2017uvr}
	N.~Roy and N.~Bhadra,
	JCAP \textbf{06} (2018), 002
	doi:10.1088/1475-7516/2018/06/002
	[arXiv:1710.05968 [gr-qc]].
	
	
	\bibitem{DeSantiago:2012nk}
	J.~De-Santiago, J.~L.~Cervantes-Cota and D.~Wands,
	Phys. Rev. D \textbf{87} (2013) no.2, 023502
	doi:10.1103/PhysRevD.87.023502
	[arXiv:1204.3631 [gr-qc]].
	
	\bibitem{Dutta:2016bbs}
	J.~Dutta, W.~Khyllep and N.~Tamanini,
	Phys. Rev. D \textbf{93} (2016) no.6, 063004
	doi:10.1103/PhysRevD.93.063004
	[arXiv:1602.06113 [gr-qc]].
	
	\bibitem{Ng:2001hs}
	S.~C.~C.~Ng, N.~J.~Nunes and F.~Rosati,
	Phys. Rev. D \textbf{64} (2001), 083510
	doi:10.1103/PhysRevD.64.083510
	[arXiv:astro-ph/0107321 [astro-ph]].
	
	\bibitem{Koivisto:2009fb}
	T.~S.~Koivisto and N.~J.~Nunes,
	Phys. Rev. D \textbf{80} (2009), 103509
	doi:10.1103/PhysRevD.80.103509
	[arXiv:0908.0920 [astro-ph.CO]].
	
	\bibitem{Bahamonde:2017ize}
	S.~Bahamonde, C.~G.~B\"ohmer, S.~Carloni, E.~J.~Copeland, W.~Fang and N.~Tamanini,
	Phys. Rept. \textbf{775-777} (2018), 1-122
	doi:10.1016/j.physrep.2018.09.001
	[arXiv:1712.03107 [gr-qc]].
	
	\bibitem{Tamanini:2014mpa}
	N.~Tamanini,
	Phys. Rev. D \textbf{89} (2014), 083521
	doi:10.1103/PhysRevD.89.083521
	[arXiv:1401.6339 [gr-qc]].
	
	\bibitem{Dutta:2017wfd}
	J.~Dutta, W.~Khyllep and N.~Tamanini,
	JCAP \textbf{01} (2018), 038
	doi:10.1088/1475-7516/2018/01/038
	[arXiv:1707.09246 [gr-qc]].

		\bibitem{Novosyadlyj:2012vd}
		B.~Novosyadlyj, O.~Sergijenko, R.~Durrer and V.~Pelykh,
		Phys. Rev. D \textbf{86} (2012), 083008
		doi:10.1103/PhysRevD.86.083008
		[arXiv:1206.5194 [astro-ph.CO]].
		
		\bibitem{Teng:2021cvy}
		Y.~P.~Teng, W.~Lee and K.~W.~Ng,
		Phys. Rev. D \textbf{104} (2021) no.8, 083519
		doi:10.1103/PhysRevD.104.083519
		[arXiv:2105.02667 [astro-ph.CO]].
		
		\bibitem{DiValentino:2020naf}
		E.~Di Valentino, A.~Mukherjee and A.~A.~Sen,
		Entropy \textbf{23} (2021) no.4, 404
		doi:10.3390/e23040404
		[arXiv:2005.12587 [astro-ph.CO]].
		
		\bibitem{Pan:2020zza}
		S.~Pan, G.~S.~Sharov and W.~Yang,
		Phys. Rev. D \textbf{101} (2020) no.10, 103533
		doi:10.1103/PhysRevD.101.103533
		[arXiv:2001.03120 [astro-ph.CO]].
		
		\bibitem{Pourtsidou:2013nha}
		A.~Pourtsidou, C.~Skordis and E.~J.~Copeland,
		Phys. Rev. D \textbf{88} (2013) no.8, 083505
		doi:10.1103/PhysRevD.88.083505
		[arXiv:1307.0458 [astro-ph.CO]].
		
		\bibitem{Sharov:2017iue}
		G.~S.~Sharov, S.~Bhattacharya, S.~Pan, R.~C.~Nunes and S.~Chakraborty,
		Mon. Not. Roy. Astron. Soc. \textbf{466} (2017) no.3, 3497-3506
		doi:10.1093/mnras/stw3358
		[arXiv:1701.00780 [gr-qc]].

\bibitem{Arkani-Hamed:2003pdi} N.~Arkani-Hamed, H.~C.~Cheng,
  M.~A.~Luty and S.~Mukohyama,
JHEP \textbf{05} (2004), 074
doi:10.1088/1126-6708/2004/05/074
[arXiv:hep-th/0312099 [hep-th]].

\bibitem{Rabiei:2015pha}
S.~W.~Rabiei, H.~Sheikhahmadi, K.~Saaidi and A.~Aghamohammadi,
Eur. Phys. J. C \textbf{76} (2016) no.2, 66
doi:10.1140/epjc/s10052-016-3907-1
[arXiv:1502.05952 [astro-ph.CO]]


\bibitem{Hinshaw:2008kr}
G.~Hinshaw \textit{et al.} [WMAP],
Astrophys. J. Suppl. \textbf{180} (2009), 225-245
doi:10.1088/0067-0049/180/2/225
[arXiv:0803.0732 [astro-ph]].
		
		
\bibitem{Ade:2013zuv}
P.~A.~R.~Ade \textit{et al.} [Planck],
Astron. Astrophys. \textbf{571} (2014), A16
doi:10.1051/0004-6361/201321591
[arXiv:1303.5076 [astro-ph.CO]].


\bibitem{Camarena:2019moy}
D.~Camarena and V.~Marra,
Phys. Rev. Res. \textbf{2} (2020) no.1, 013028
doi:10.1103/PhysRevResearch.2.013028
[arXiv:1906.11814 [astro-ph.CO]]




\end{thebibliography}
\end{document}